\documentclass{PoS}

\title{Vacuum Stability in two-Higgs doublet models}

\ShortTitle{Vacuum Stability in two-Higgs doublet models}

\author{Nuno Barros e S\'a \\
        DCTD, Universidade dos A\c cores\\
        9500-801 Ponta Delgada, Portugal\\
        E-mail: \email{nunosa@uac.pt}}

\author{Augusto Barroso \\
        Centro de F\'{\i}sica Te\'orica e Computacional, Faculdade
de Ci\^encias, Universidade de Lisboa,\\ Avenida Professor Gama
Pinto, 2, 1649-003 Lisboa, Portugal; \\
        E-mail: \email{barroso@cii.fc.ul.pt}}

\author{Pedro Ferreira \\
Instituto Superior de Engenharia de Lisboa, Rua
Conselheiro Em\'{\i}dio Navarro 1, 1959-007 Lisboa, Portugal;\\
Centro de F\'{\i}sica Te\'orica e Computacional, Faculdade de
Ci\^encias, Universidade de Lisboa,\\ Avenida Professor Gama
Pinto, 2, 1649-003
Lisboa, Portugal. \\
        E-mail: \email{ferreira@cii.fc.ul.pt}}

\author{\speaker{Rui Santos}%
         \thanks{This work is supported by Funda\c{c}\~ao para a
Ci\^encia e Tecnologia under contract PTDC/FIS/70156/2006. R.S. is
supported by the FP7 via a Marie Curie Intra European Fellowship,
contract number PIEF-GA-2008-221707. A special thanks to Per Osland.}\\
        NExT Institute and School of Physics and Astronomy,
University of Southampton Highfield, Southampton SO17 1BJ, United
Kingdom\\
        E-mail: \email{santos@pp.rhul.ac.uk}}

\abstract{In this work we review the status of tree-level vacuum
stability in general two-Higgs doublet models. We also discuss the
problem of Normal minima in some classes of potentials. In some of
these potentials, Normal minima can coexist leading to potentially
dangerous physic scenarios as, for instance, a two step
spontaneous symmetry breaking.}

\FullConference{Prospects for Charged Higgs Discovery at Colliders\\
                 16-19 September 2008\\
                 Uppsala, Sweden}

\begin{document}

\section{Introduction}

The Standard Model (SM) is a theory invariant under the symmetry
group $SU(3)_c \otimes SU(2)_L \otimes U(1)_Y$. In the SM, the
spontaneous symmetry breaking (SSB) mechanism is accomplished with
one scalar doublet whose vacuum expectation value (vev) is usually
written as $<\phi_{SM}> = (0 \quad \,   v/\sqrt{2})$ where $v=246$
GeV. The charge operator is defined as $Q=1/2 (\sigma_3 + Y I)$,
where $\sigma_3$ is the standard Pauli matrix, $I$ is the $2
\times 2$ identity matrix and $Y$ is the weak hypercharge. With
this definition $Q \, <\phi_{SM}> = 0$ as it should because the
vacuum has no electric charge. Had we started with the most
general vacuum configuration $<\phi_{SM}> =(v_1+iv_2 \quad \,
v_3+iv_4)/\sqrt{2}$ we would still have a massless photon. The
electromagnetic $U(1)_{em}$ can not be broken with only one
doublet and to choose the most general vacuum configuration simply
amounts to a redefinition of the charge operator.

The situation changes radically once one adds a second doublet. We
have now eight fields that can acquire a vev.  We can however use
the $SU(2)_L \otimes U(1)_Y$ gauge freedom to write the most
general vacuum structure as
\begin{equation}
\left(
\begin{array}{c}  \, 0 \, \\ \, v_1 \, e^{i\theta_1} \,
\end{array}
\right) \qquad \quad \left(
\begin{array}{c}  \, v_2 \, \\ \, v_3 \,
\end{array}
\right) \, \, .
\end{equation}
The gauge boson's mass matrix can now have four non-zero
eigenvalues. The mass eigenvalue related to the photon is given by
\begin{equation}
m_{\gamma}^2= \frac{1}{8} \left[ v^2 (g^2+g'^2)-
\sqrt{v^4(g^2+g'^2)-16 g^2 \, g'^2 \, v_1^2\, v_2^2} \right]
\end{equation}
where $v=\sqrt{v_1^2+v_2^2+v_3^2}$ and $g$ and $g'$ are the
$SU(2)$ and $U(1)$ gauge couplings respectively. There are two
ways to recover a massless photon: either by setting $v_1=0$ (the
SM case) or by choosing $v_2=0$ and the vevs are then
aligned~\cite{DiazCruz:1992uw}. Otherwise the photon becomes
massive as a consequence of the charged vacuum configuration.

Gauge invariance disallows not only a charge breaking vacuum but
also a CP breaking vacuum in the SM. Again the situation changes
in THDM. Defining a CP transformation as $\phi_i \rightarrow
\phi_i^*$, three different types of vacua can be defined for a
general THDM with all constants real,
\begin{equation}
<\phi_1>_N = \left(
\begin{array}{c}  \, 0 \, \\ \, v_1 \,
\end{array}
\right) \qquad \quad <\phi_2>_N = \left(
\begin{array}{c}  \, 0 \, \\ \, v_2 \,
\end{array}
\right) \, \, ,
\end{equation}
which we call the normal vacuum,
\begin{equation}
<\phi_1>_{CB} = \left(
\begin{array}{c}  \, 0 \, \\ \, v^\prime_1 \,
\end{array}
\right) \qquad \quad <\phi_2>_{CB} = \left(
\begin{array}{c}  \, \alpha \, \\ \, v^\prime_2 \,
\end{array}
\right) \, \, ,
\end{equation}
for a vacuum that breaks charge conservation, and finally
\begin{equation}
<\phi_1>_{CP} = \left(
\begin{array}{c}  \, 0 \, \\ \, v^{\prime\prime}_1\,+\,i\delta \,
\end{array}
\right) \qquad \quad <\phi_2>_{CP} = \left(
\begin{array}{c}  \, 0 \, \\ \, v^{\prime\prime}_2 \,
\end{array}
\right) \, \, ,
\end{equation}\label{eq:CPvac}for a CP breaking vacuum.
Vacua with $\alpha$ and $\delta$ simultaneously non-zero are not
considered because the minimisation conditions of the potential
forbid them~\footnote{Except for a very special case in the
explicit CP breaking potential. Even in that case, though, via a
basis change, that vacuum may be reduced to one with
$\alpha\,\neq\,0$ and $\delta\,=\,0$.}.
\section{The THDM potentials}

The vacuum structure of THDM has been the subject of many
studies~[1-11], while the vacuum structure of an arbitrary number
of doublets was studied in~\cite{Barroso:2006pa, Nishi:2006tg}.
The scalar sector of a THDM is built with eight independent
fields, four from each doublet. The number of independent gauge
invariants is however four due to the SM gauge invariance under
$SU(2)_L \otimes U(1)_Y$ with four gauge generators~\footnote{A
similar counting for the SM with one doublet would naively lead to
the conclusion that we would end up with zero gauge invariants.
However this reasoning does not apply to the SM due to the
existence of a little group, subgroup of the gauge group, that
leaves the doublet invariant.}. Therefore, all properties of the
potential can be studied in terms of four independent gauge
invariants, which we choose to be $x_1 = |\phi_1|^2$, $x_2 =
|\phi_2|^2$, $x_3 = Re(\phi_1^\dagger\phi_2)$ and $x_4 =
Im(\phi_1^\dagger\phi_2)$. In terms of this basis set of gauge
invariants, the most general renormalizable THDM has 14 real
parameters and can be written as
\begin{eqnarray}
V \;\;=& \;\; a_1\, x_1\, + \,a_2\, x_2\, + \,a_3 x_3 \,+\, a_4
x_4 \,+\, b_{11} \, x_1^2\, +\, b_{22}\, x_2^2\, +\, b_{33}\,
x_3^2\, +\, b_{44}\, x_4^2\, +\,
\nonumber \\
 & \;\; b_{12}\, x_1 x_2\, +\, b_{13}\, x_1 x_3\, + b_{14}\,
x_1 x_4\, +\,b_{23}\, x_2 x_3 +\,b_{24}\, x_2 x_4 +\,b_{34}\, x_3
x_4\;\; . \label{eq:pot14}
\end{eqnarray}
Under the CP transformation of the form $\phi_i \rightarrow
\phi_i^*$, $x_1$, $x_2$ and $x_3$ remain the same but $x_4$
switches signal. Thus, the terms of the potential which are linear
in $x_4$ break CP explicitly. Therefore a potential for which CP
is not explicitly broken cannot have terms linear in $x_4$ in
eq.~\ref{eq:pot14}: $a_4 = b_{14} = b_{24} = b_{34} = 0$. This
potential has 10 free parameters and can still break CP
spontaneously.

To further reduce the number of parameters while keeping the model
renormalizable we can force the potential to be invariant under
certain symmetries imposed to the fields. In the case of THDM the
$Z_2$ and a softly broken global $U(1)$ symmetries
lead~\cite{Velhinho:1994vh} to two seven parameter models with a
very different phenomenology. Recently, it was shown
in~\cite{Ferreira:2008zy} that these are the only two "simple"
symmetries that can be imposed on THDM. In~\cite{Velhinho:1994vh}
an interesting relation between CP-violation, symmetries and
flavour changing neutral currents (FCNC) at tree level was found.
In fact, for the 10-parameter potential only one of the
CP-breaking stationary conditions can be forced to have no
solution. This equation can be written as
\begin{equation}
b_{13} \, (v^{\prime\prime \, 2}_1 + \delta^2)\, + b_{23} \,
v^{\prime\prime \, 2}_2 + (b_{33}-b_{44}) \, v^{\prime\prime}_1
v^{\prime\prime}_2 \, = \, a_3
\end{equation}
and renormalizability forces $b_{13}$ and $b_{23}$ to be
simultaneously either zero or non-zero. Therefore the equation has
no solution in the two following situations
\begin{eqnarray}
& & b_{13} = b_{23} = a_3 = 0; \, \, \,    b_{33} \neq b_{44}
\qquad (\phi \rightarrow - \phi_1 \, ; \qquad \phi_2 \rightarrow
\phi_2)
\nonumber \\
& & b_{13} = b_{23} = b_{33}-b_{44} = 0; \, \, \,    a_{3} \neq 0
\qquad (\phi \rightarrow e^{i \theta} \phi_1 \, ; \qquad \phi_2
\rightarrow \phi_2).
\end{eqnarray}
Hence, forcing the CP-breaking minimum not to exist leads to two
phenomenologically different 7-parameter potentials. Furthermore,
by extending the symmetry to the Yukawa sector one avoids the
existence of FCNC at tree level. There are still two other
potentials to consider: the 6-parameter potential with an axion
(exact $U(1)$ symmetry) or a potential that softly breaks $Z_2$
which has 8 free parameters and can be written as
\begin{equation}
V_{Z_2}^{soft} =  a_1\, x_1\, + \,a_2\, x_2  + \,a_3\, x_3 \,+\,
b_{11} \, x_1^2\, +\, b_{22}\, x_2^2\, +\, b_{33}\, x_3^2\, +\,
b_{44}\, x_4^2\, +\, b_{12}\, x_1 x_2 \;\; .\label{eq:potZ2S}
\end{equation}
From the phenomenological point of view all these models can be
studied as a limiting case of $V_{Z_2}^{soft}$ in the case of
normal minima. Choosing as free parameters of $V_{Z_2}^{soft}$ the
four Higgs masses, $\tan \beta = v_2/v_1$, $\alpha$ (the rotation
angle in the CP-even sector) and $a_3$, if we set $a_3=0$ we
obtain the potential with the exact $Z_2$ symmetry, $V_{Z_2}$ and
when $a_3= -M_A^2/sin(2 \beta)$ we find the potential with the
softly broken $U(1)$ symmetry, $V_{U(1)}^{soft}$.
\section{Minima of different nature}

Contrary to the SM, THDM can have several stationary points of
different natures. The three possible types of stationary points
which we called Normal (N), Charge Breaking (CB) or CP-breaking
(CP) were defined in the introduction. Therefore, one may inquire
about the possibility of having two minima of different nature for
the same set of parameters. This could pose a problem to THDM as
we could be living in a Normal minimum and suddenly tunnel to a
deeper charge breaking minimum where the photon acquires a mass.
To answer this question we have to compare the vacuum energy at
the different stationary points, $V_{CB}$, $V_{CP}$ and $V_{N}$.
In \cite{Ferreira:2004yd} we found a very interesting relation
between the difference of the value of the potential at a CB
stationary point, $V_{CB}$, and the value of the potential at a
normal stationary point, $V_N$,
\begin{equation}
V_{CB}\;-\;V_N\;\; = \;\; \frac{M^2_{H^\pm}}{2\,v^2} \;\left[
(v^\prime_1\,v_2 \;-\;v^\prime_2\,v_1)^2\; + \;
\alpha^2\,v_1^2\right]\;\;\; , \label{eq:difcb}
\end{equation}
where $v^2\,=\,v_1^2\,+\,v_2^2$ and $M^2_{H^\pm}$ is the value of
the squared mass of the charged Higgs scalar, evaluated at the
normal stationary point. The normal stationary point is a minimum
if all squared scalar masses are positive. Hence, this equation
tells us that if the normal stationary point is a minimum, it is
definitely below the charge breaking stationary point.
Furthermore, in ref.~\cite{Ferreira:2004yd} we proved that in this
case the charge breaking stationary point is a saddle point. The
stability of the normal minimum against tunneling to a deeper
charge breaking stationary point is thus ensured in THDM.

A similar result holds when one compares a CP and a normal
stationary point. When CP is a good quantum number at the
potential level, we found that the difference between the value of
the potential at the CP stationary point, $V_{CP}$, and at the
normal stationary point, $V_N$,is given by~\cite{Ferreira:2004yd}
\begin{equation}
V_{CP}\;-\;V_N\;\; = \;\; \frac{M^2_A}{2\,v^2} \;\left[
(v^{\prime\prime}_1\,v_2 \;-\;v^{\prime\prime}_2\,v_1)^2\; + \;
\delta^2\,v_2^2\right]\;\;\; . \label{eq:difcp}
\end{equation}
$M^2_A$ is now the value of the squared pseudoscalar mass at the
normal stationary point. Again, if the normal stationary point is
a minimum, positivity of $M^2_A$ ensures that the CP stationary
point is above it. Furthermore, I. Ivanov has
proved~\cite{Ivanov:2007de} that in this situation, the CP
stationary is a saddle point. As the CP stationary point is also
uniquely determined, the stability of the normal minimum against
tunneling is guaranteed.
Similar results hold for the CP breaking and for the CB minima. If
a CP breaking stationary point is a minimum, the competing normal
and charge breaking stationary points are saddle points above it -
the CP breaking minimum is then a global one. Although probably
not of fundamental importance, the CB stationary point, when a
minimum is the global one.
\section{Normal minima}

There is still one situation that deserves our attention - the
case of two simultaneous normal minima. The THDM can have at most
two normal minima~\cite{Ivanov:2007de} and from last section we
know that minima of different nature never coexist. However we
could still have two minima that had different spontaneous
symmetry breaking patterns with different masses for the gauge
bosons. In the most general 14-parameter potential with explicit
CP-violation the relation between two normal stationary points,
$N_1$ and $N_2$ as defined in eq. (1.5) is given by
\begin{equation}
V_{N_2}\;-\;V_{N_1} \;\; = \;\;\frac{1}{2}\left[
\left(\frac{M^2_{H^\pm}}{v^2}\right)_{N_1}\;-\;\left(\frac{M^2_{H^\pm}}{v^2}
\right)_{N_2} \right] \;\left[(v^{\prime\prime}_1\,v_2 \;-\;
v^{\prime\prime}_2\,v_1)^2\; + \; \delta^2\,v_2^2\right]\;\;\; .
\label{eq:difn12}
\end{equation}
In this equation we have $(v^2)_{N_1}\,=\,v_1^2\,+\,v_2^2$ and
$(v^2)_{N_2}\,=\,{v^{\prime\prime}_1}^2\,+\,{v^{\prime\prime}_2}^2\,+\,\delta^2$,
and $(M^2_{H^\pm})_{N_{1,2}}$ are the squared charged scalar
masses at each of the $N_1$, $N_2$ stationary points. This
interesting relation tells us that the deepest stationary point
will be the one with the largest ratio between the square of the
charged Higgs mass and the $v^2$. This means that the deeper
minimum has the largest splitting between the charged Higgs mass
and the "theoretical" $W$ boson mass. In a CP conserving
potential, the equation is similar but with $\delta=0$ because the
normal minima configuration have no phases in a CP conserving
potential. However this expression adds very little to the problem
of the competing normal minima.

We have mentioned that the CB and CP stationary points are unique
since they are given by linear equations on the vevs. However,
this is not true for the normal stationary points. The
stationarity conditions are always a set of two coupled cubic
equations which can only be solved analytically for the potential
with the exact $Z_2$ symmetry, $V_{Z_2}$, where the stationarity
points are again uniquely determined. Therefore the stationarity
equations can only be solved numerically. The question we are
addressing now is the following: is it possible to have a normal
minimum for a definite set of parameters with the right gauge
boson masses and have another normal minimum below it with
completely different masses for the gauge bosons? And the answer
to this question is yes as long as the soft breaking term is
present. As an example we can have for $V_{Z_2}^{soft}$ a local
minimum with $m_H = m_{H^\pm} = m_A = 300$ GeV, $m_h=100$ GeV and
$m_w = 80.4$ GeV and a global minimum with $m_H = m_A = 436$ GeV
$m_{H^\pm} = 365$ GeV, $m_h = 190$ GeV and $m_w = 107.5$ GeV. The
vacuum energy difference between the two is $V_G - V_L = -4.2
\times 10^{8}$. Hence, tunneling effects can occur in this
situation and the subject needs further and more detailed study.
Note however that all potentials are well behaved for the CP
breaking minima no matter how small the CP violation phase is
because the CP breaking minima is uniquely determined. Finaly,
only the potential with the exact $Z_2$ symmetry guarantees that
the minimum, being a normal one, is unique. Even the competing
normal stationary points with one of the vevs set to zero are
above it. For a discussion on this point
see~\cite{Barroso:2007rr}.

\section{Conclusions}
\noindent
In this section we try to sum up our present knowledge about the
tree-level vacuum of THDM in the following
points~\cite{Velhinho:1994vh, Ferreira:2004yd, Ivanov:2006yq,
Barroso:2007rr, Ivanov:2007de}:
\begin{itemize}

\item THDM bounded from below have no maxima (except for the
trivial);

\item THDM have at most two minima;

\item Minima of different nature never coexist;

\item Unlike Normal minima, CB and CP minima are uniquely
determined;

\item If a THDM has only one normal minimum then this is the
absolute minimum - all other stationary points if they exist are
saddle points and above it;

\item If a THDM has a CP breaking minimum then this is the
absolute minimum - all other stationary points if they exist are
saddle points and above it;

\item THDM with no explicit CP breaking show problems for two
competing normal minima;

\item The THDM with an exact $Z_2$ symmetry has several
interesting features: the minimization equations are uniquely
determined which is not true when the soft breaking symmetry term
is present; being in a normal minimum \textit{automatically}
guarantees tree-level vacuum stability - no extra conditions have
to be imposed; it is the only THDM where the only masses present
in all scalar couplings are the ones in the corresponding
interaction vertex - again this is no longer true when the soft
breaking term is present. It has however the drawback of not
allowing for CP violating minima. This version of the THDM is not
related to the MSSM, where the existence of the soft breaking term
is mandatory to avoid the existence of an axion.

\end{itemize}


\begin{thebibliography}{99}


\bibitem{DiazCruz:1992uw}
  J.~L.~Diaz-Cruz and A.~Mendez,
  Nucl.\ Phys.\  B {\bf 380} (1992) 39.

\bibitem{Velhinho:1994vh}
  J.~Velhinho, R.~Santos and A.~Barroso,
  Phys.\ Lett.\  B {\bf 322} (1994) 213.

\bibitem{Ferreira:2004yd}
  P.~M.~Ferreira, R.~Santos and A.~Barroso,
  Phys.\ Lett.\  B {\bf 603} (2004) 219, [Erratum-ibid.\  B {\bf 629} (2005)
  114]; A.~Barroso, P.~M.~Ferreira and R.~Santos,
  Phys.\ Lett.\  B {\bf 632} (2006) 684.

\bibitem{Ginzburg:2004vp}
  I.~F.~Ginzburg and M.~Krawczyk,
  Phys.\ Rev.\  D {\bf 72} (2005) 115013.

\bibitem{Barroso:2005tq}
  A.~Barroso, P.~M.~Ferreira and R.~Santos,
  Afr.\ J.\ Math.\ Phys.\  {\bf 3} (2006) 103.

\bibitem{Gunion:2005ja}
   G.~C.~Branco, M.~N.~Rebelo and J.~I.~Silva-Marcos,
   Phys.\ Lett.\  B {\bf 614} (2005) 187;
  J.~F.~Gunion and H.~E.~Haber,
  Phys.\ Rev.\  D {\bf 72} (2005) 095002.

\bibitem{Maniatis:2006fs}
  M.~Maniatis, A.~von Manteuffel, O.~Nachtmann and F.~Nagel,
  Eur.\ Phys.\ J.\  C {\bf 48} (2006) 805.

\bibitem{Ivanov:2006yq}
  I.~P.~Ivanov,
  Phys.\ Rev.\  D {\bf 75} (2007) 035001, [Erratum-ibid.\  D {\bf 76} (2007)
  039902].

\bibitem{Ginzburg:2007jn}
  I.~F.~Ginzburg and K.~A.~Kanishev,
  Phys.\ Rev.\  D {\bf 76} (2007) 095013.

\bibitem{Barroso:2007rr}
  A.~Barroso, P.~M.~Ferreira and R.~Santos,
  Phys.\ Lett.\  B {\bf 652} (2007) 181.

\bibitem{Ivanov:2007de}
  I.~P.~Ivanov,
  Phys.\ Rev.\  D {\bf 77} (2008) 015017.

\bibitem{Barroso:2006pa}
  A.~Barroso, P.~M.~Ferreira, R.~Santos and J.~P.~Silva,
  Phys.\ Rev.\  D {\bf 74} (2006) 085016.

\bibitem{Nishi:2006tg}
  C.~C.~Nishi,
  Phys.\ Rev.\  D {\bf 74} (2006) 036003, [Erratum-ibid.\  D {\bf 76} (2007)
  119901];
  C.~C.~Nishi,
  Phys.\ Rev.\  D {\bf 76} (2007) 055013.

\bibitem{Ferreira:2008zy}
  P.~M.~Ferreira and J.~P.~Silva, arXiv:0809.2788 [hep-ph], to appear in PRD.



\end{thebibliography}
\end{document}